\documentclass[aps,english,nofootinbib,twocolumn]{revtex4-1}
\usepackage[T1]{fontenc}
\usepackage[utf8]{inputenc}
\usepackage{color}
\usepackage{babel}
\usepackage{pmboxdraw}
\usepackage{amstext}
\usepackage{graphicx}
\usepackage{natbib}
\usepackage{float}
\usepackage{url}  
\usepackage{booktabs}  
\usepackage{amsfonts}  
\usepackage{nicefrac}  
\usepackage{cleveref}
\usepackage{float}

\usepackage[none]{hyphenat} 

\begin{document}

\title{Electronic Phase Transformations and Energy Gap Variations in Uniaxial and Biaxial Strained Monolayer VS$_2$ TMDs: A Comprehensive DFT and Beyond-DFT Study}
  
\author{Oguzhan Orhan}
\affiliation{Faculty of Science, Department of Physics, Firat University, Elazig 23119, Turkey}
\email{oguzhan.orhan@firat.edu.tr}

\author{Şener Özönder}
\affiliation{Institute for Data Science \& Artificial Intelligence, Bogazici University, İstanbul 34342, Turkey}

\author{Soner Ozgen}
\affiliation{Faculty of Science, Department of Physics, Firat University, Elazig 23119, Turkey}

\begin{abstract}
In the field of 2D materials, transition metal dichalcogenides (TMDs) are gaining attention for electronic applications. Our study delves into the H-phase monolayer VS$_2$ of the TMD family, analyzing its electronic structure and how strain affects its band structure using Density Functional Theory (DFT). Using a variety of computational methods, we provide an in-depth view of the electronic band structure. We find that strains between -5\% and +5\% significantly affect the energy gap, with uniaxial strains having a stronger effect than biaxial strains. Remarkably, compressive strains induce a phase shift from semiconducting to metallic, associated with symmetry breaking and changes in bond length. These findings not only deepen our understanding of the electronic nuances of monolayer VS$_2$ under varying strains but also suggest potential avenues for creating new electronic devices through strain engineering.
\end{abstract}


\maketitle


\section{Introduction}
\par Transition metal dichalcogenides (TMDs) are a family of two-dimensional (2D) materials that comprise a sandwich-like structure, X-M-X, where X is a chalcogen atom (S, Se, or Te) and M is a metal atom (Mo, W, V, Nb, etc.). Each layer is stacked vertically through van der Waals interactions.  The metal atoms in TMDs are coordinated in a trigonal prismatic (H) or octahedral (T) configuration \cite{1,2,3}. They exhibit various electronic properties such as superconductors, metals, insulators, or topological materials depending on the coordination type and the X and M atoms. These electronic properties of TMDs, which enable functions ranging from controlling electron flow based on spin and valley to creating light-emitting devices, manufacturing transistors and switches, and advancing biomedical technologies, are notably influenced by the phase transition they undergo under specific stimuli \cite{2,4,5,6,7,8}.
\par TMDs exhibit electronic phase transitions due to charge doping, vacancy, temperature, and electric field \cite{2,9,10,11,12}. Also, the number of layers in TMDs is directly related to their electronic properties, previous ab initio calculations have reported that the absence of inversion symmetry directly influences their band structures \cite{13,14,15}. Furthermore, strain engineering is a technique that involves applying strain to alter the electronic properties of materials and has been demonstrated in both experimental studies and ab initio calculations. For instance, multilayer MoS$_2$ undergoes the electronic transition from semiconductor to metallic state at ~19 GPa \cite{16}, and MoTe$_2$ is altered from a 2H phase with semiconductor properties to a 1T$^\prime$ phase with metallic characteristics by introducing a tensile strain of 0.2\% \cite{17}. Strain induces ferroelectric phase transformation and direct to indirect band evolution, according to experimental \cite{18} and theoretical studies \cite{19}. However, despite the significant potential and diverse electronic properties of TMDs, there is a lack of comprehensive understanding of the effect of strain, particularly uniaxial strain, on the electronic structure of specific TMDs such as monolayer H-phase VS$_2$. Furthermore, the existing literature predominantly focuses on Group 6 TMDs, neglecting the exploration of the unique characteristics of other TMDs, particularly those based on Group 5 elements like VS$_2$. The paucity of research in this area hampers the expansion of potential applications for these materials.
\par Vanadium-based Group 5 transition metal dichalcogenides (TMDs) exhibit a fascinating interplay between electronic properties and crystal structure. Monolayer vanadium disulfide (VS$_2$) exemplifies this relationship, existing in two distinct polymorphs: hexagonal (H) and trigonal (T) phases \cite{9}. Theoretical studies have established that the H phase is a narrow-band gap ferromagnet \cite{20,21,22}, but T-phase has a metallic character while retaining magnetism.\cite{20,23,24}. Despite the H phase's predicted thermodynamic stability, the energy difference between these structures remains remarkably small (65 meV) \cite{24}. Recent experimental investigations have corroborated these theoretical predictions, confirming the presence of charge density waves in the T \cite{25,26} phase and p-type semiconductivity in the H phase \cite{27}. Furthermore, ab initio calculations suggest that the electronic, optical, and magnetic properties of VS$_2$ can be significantly modulated by factors such as the number of layers, charge doping, and elemental substitution \cite{13,20,21}.
\par The influence of strain on the electronic structure of VS$_2$, particularly the H phase, has garnered increasing attention due to its potential to unlock novel functionalities. Biaxial strain has emerged as a powerful tool for manipulating its properties, with studies exploring its impact on phase stability and electronic transitions. Kan et al. observed an insulator-to-metal transition in the H phase under compressive biaxial strain (-3\% to -4\%), followed by a bandgap widening with increasing tensile strain (up to 0\%) \cite{22}. Similarly, Fuh et al. reported that the H phase becomes metallic at -4\% and -2\% strain, with the bandgap affected by tensile strain up to 6\% before transitioning to metal at higher strains (8\% and 10\%)  \cite{23}. Tariq et al. investigated this phenomenon and found that the H phase becomes metallic at -3\% strain. Additionally, they observed a decreasing bandgap with increasing biaxial strain (-2\% to 3\%). \cite{tariq2021pristine}. Although prior research has primarily focused on biaxial strain effects on VS$_2$'s H phase, the implications of uniaxial strain are less understood, yet crucial for applications in heterojunctions and straintronics \cite{24,28}. For instance, Kim et al. have considered the effect of strain due to lattice mismatch in the formation of graphene-VS$_2$ heterostructures \cite{kim2021electronic}. Our study aims to address the limitation by investigating the effects of both uniaxial and biaxial strain on the H-phase of monolayer VS$_2$. The goal is to enhance the understanding of strain mechanisms for advanced materials and device development.

\par In this study, we utilize DFT to elucidate the electronic structure of monolayer VS$_2$ in its H-phase. We have critically contrasted two prominent pseudopotential (PP) libraries for VS$_2$. Our methodology incorporates advanced techniques such as semi-empirical DFT+U, GAUPBE hybrid function, G0W0, and self-consistent GW methods, aiming for a comprehensive exploration of the VS$_2$ band gap. The main focus of this research is to determine the impact of strain on the band structure of monolayer VS$_2$. We have subsequently examined its band structure under both uniaxial and biaxial strains of up to $\pm~5\%$ using two PP approximations. The results yielded a notable observation: Monolayer VS$_2$ undergoes an electronic phase transition from a semiconductor to a metal under compressive strain. Conversely, applying tensile strain consistently widened the band gap as the strain ratio increased.

\section{Computational Details}

\par In our research, we assessed the electronic band structure of monolayer H-phase VS$_2$, particularly under strain. We applied both uniaxial and biaxial strains, ranging from -5\% to +5\%, to investigate strain-induced electronic phase transitions. The lattice deformations were generated using the Pymatgen software \cite{29}. DFT calculations in this study utilized QUANTUM ESPRESSO (QE) software \cite{30}. We compared two pseudopotential libraries: the SSSP's precision version \cite{31}, which includes ultrasoft (US) PP \cite{32,33} for its efficiency and lower energy cutoff, and the Optimized Norm-Conserving Vanderbilt (ONCV) PP \cite{34, schlipf2015optimization}, selected for its high precision but with increased computational cost. This strategic choice of pseudopotentials, focusing on the balance between computational efficiency and detailed electronic structure analysis, is crucial for understanding the behavior of VS$_2$ under strain. The study utilized the generalized gradient approximation (GGA) of the Perdew, Burke, and Ernzerhof (PBE) exchange-correlation functional \cite{35}, with calculations based on the ferromagnetic order preference of VS$_2$'s H phase \cite{20}.
\par The energy cutoff of plane-wave basis expansion for ONCV and US were set at 80 Ry and 40 Ry, respectively. The energy and force convergence criteria for both optimized and relaxed structures were 1 meV and 1 meV/\AA, respectively, with a $12~\times~12~\times~1$ gamma-centered k-point mesh. During the optimization of the deformed structures, the lattice vectors remained fixed while the atoms were relaxed. For self-consistent and non-self-consistent steps, the k-point mesh was set to $16~\times~16~\times~1$ and $24~\times~24~\times 1$, respectively. To avoid the periodic boundary condition, we used a vacuum space of 15 \AA~along the z-axis.
\par Various computational methods, including DFT+U, the GAUPBE hybrid function \cite{36} G0W0, and the self-GW approach, were employed to accurately determine the energy band gap of monolayer VS$_2$. The Hubbard parameter for the V 3d orbital in DFT+U calculations was chosen as 4.1 eV \cite{37}. Given its superior computational cost and convergence with a smaller q mesh, the GAUPBE function was preferred over HSE0 \cite{36}. The q-point sampling for the hybrid function was selected as $6~\times~6~\times~1$. Since obtaining the band structure from the hybrid function in QE can be problematic, we adopted the Wannier90 code,  which implements the Wannier interpolation \cite{38}.
\par G0W0 and self-GW calculations based on the iterative convergence of G and W concerning the eigenenergies of G0W0 were performed with the Yambo package \cite{39}. We set up the Brillouin Zone sample via a $16 \times 16 \times 1$ k grid. The plasmon pole approximation was utilized to solve for the dielectric function $\varepsilon_{G-G^{'}}(\omega-q)$ \cite{40,41}. A cutoff of 8 Ry was used for the size of the dielectric matrix, including up to 420 states in the sum-over-state of the response function. The same number of states was used in the calculation of the correlation part of the self-energy. The random integration method was also employed to achieve band convergence at lower k points \cite{42}.

\section{Results and Discussion}
\subsection{Equilibrium Lattice Parameters}

\par
\begin{figure}[t]
\centering
  \includegraphics[width=8.2cm, height=7.0cm]{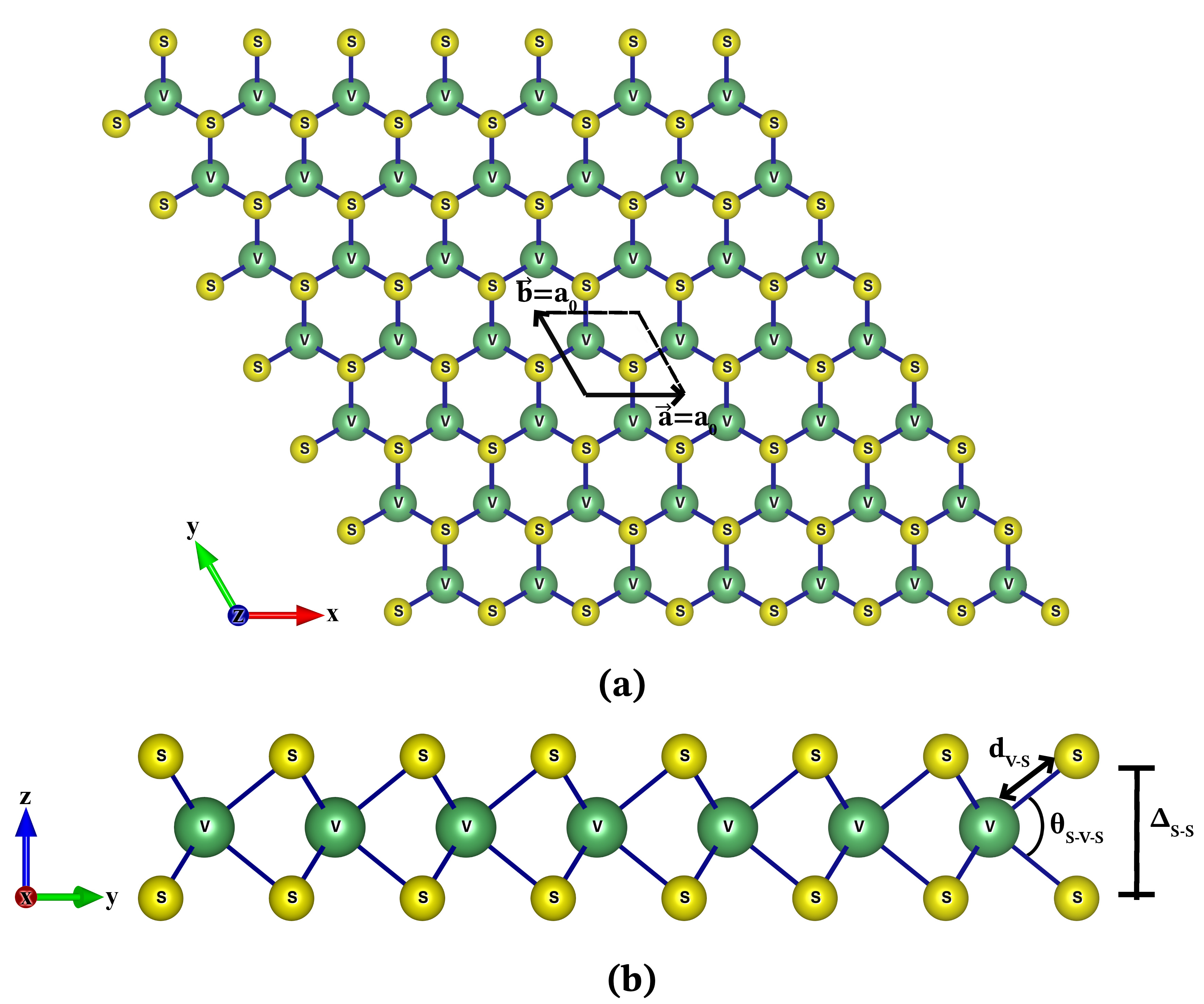}
  \caption{(a) Lattice constant a$_0$ with the top view, (b) the V-S bond length d$_{(V-S)}$, the buckling height between the two S atom planes $\Delta_{S-S}$, and the V-S-V bond angle $\theta_{S-V-S}$ with the side view.}
  \label{fgr:figure1}
\end{figure}

 Since experimental studies of monolayer VS$_2$ are relatively limited, both ONCV and US PPs have been employed to attempt to better understand its structure. Table~\ref{tbl:table1} includes the optimized geometrical parameters of VS$_2$ using both ONCV and US PPs, offering a comparative perspective compared with previous studies. Figure~\ref{fgr:figure1} illustrates the geometric structures of the H phase of monolayer VS$_2$, with Figure~\ref{fgr:figure1} (a) and (b) showing the top and side views, respectively.
 
\par These parameters are fundamental to understanding the inherent structural properties of monolayer VS$_2$. The ONCV PP yields a lattice constant of 3.18~\AA, a bond length of 2.365~\AA, a buckled height of 2.982~\AA, and a bond angle of 78.16$^{\circ}$. Comparable values are obtained with the US PP: 3.173~\AA, 2.361~\AA, 2.98~\AA, and 78.249$^{\circ}$, respectively. Furthermore, PAW PP exhibits fairly consistent values with both of them. Indeed, despite the use of different pseudopotentials (ONCV, US in this study and PAW in previous studies), the differences in lattice properties are minimal, confirming the consistency and validity of the comparative analysis. All three PP approximations are in agreement with the experiment result \cite{27}. After optimizing the lattice properties of monolayer VS$_2$, the electronic band structure has been computed using these PPs and the several methods they allow.

 \begin{table}[!htbp]
    \caption{Optimized geometrical parameters of H-phase VS$_2$ and their comparison with previous studies: Lattice parameter a$_0$; Bond length V-S (d$_{V-S}$); Buckled height between two S atom planes ($\Delta_{S-S}$); V-S-V bond angle, $\theta_{V-S-V}$.}
    \label{tbl:table1}
    \begin{tabular*}{0.48\textwidth}{@{\extracolsep{\fill}}p{0.8cm} p{1.7cm} p{1.5cm} p{1.4cm} p{1.0cm}}
        \toprule
        PP & a$_0$({\AA}) & d$_{V-S}$({\AA}) & $\Delta_{S-S}$({\AA}) & $\theta_{V-S-V}$ \\
        \midrule
        ONCV& 3.180& 2.365 &	2.982 &	84.484 \\
        US  & 3.173 & 2.361 &	2.980 &	84.370 \\
        PAW & 3.173 \cite{22}\newline 3.174 \cite{43} & 2.362 \cite{23}\newline 1.469 \cite{43} & 2.982 \cite{23} & 84 \cite{22} \\
        \bottomrule
    \end{tabular*}
\end{table}

\subsection{Density of States and Electronic Band Structure}

\par We use two different pseudopotential approximations and a variety of computational methods, including DFT+U, GAUPBE, G0W0, and self-GW, to provide a detailed insight into its electronic band structure. Table~\ref{tbl:table2} provides a comprehensive comparison of the band gap results for monolayer VS$_2$, categorized by PPs, method, gap value, transition point, and spin channel.

\begin{table}[!b]
  \caption{Comparative band gap energies of H-phase monolayer VS$_2$ and symmetry points and spin channels where the gap comes about.}
  \label{tbl:table2}
  \begin{tabular*}{0.48\textwidth}{@{\extracolsep{\fill}}p{0.8cm} p{1.1cm} p{1.4cm} p{1.5cm} p{1.5cm}}
    \toprule 
    PP & Method & Gap (eV) & Transition Point & Spin\newline Channel \\
    \midrule
         &  PBE     &   0.129   &   G\(\rightarrow\)M  &   up-down  \\
    ONCV &  G0W0    &   0.987   &   G\(\rightarrow\)M  &   up-down  \\
         &  Self$-$GW &   1.610   &   K\(\rightarrow\)M  &   up-down  \\
         &          &           &                      &            \\
     US  &  PBE	   &   0.018   &   G\(\rightarrow\)M  &   up-down  \\
         &  DFT+U   &   0.617   &   G\(\rightarrow\)K  &   up-up    \\
         &  GAUPBE  &   0.946   &   G\(\rightarrow\)K  &   up-up    \\
         &          &           &                      &            \\
         &  PBE     & 0.187 \cite{22} \newline 0.046 \cite{23} & G\(\rightarrow\)M \newline \cite{22,23} & up-down \cite{22,23} \\
         &  DFT+U   & 0.26 \cite{20} \newline 0.473 \cite{23} & G\(\rightarrow\)M \newline \cite{20,23} & up-down \cite{20,23} \\
    PAW  &          & 0.721 \cite{22}  & G\(\rightarrow\)K \cite{22} & up-up \cite{22}  \\
         &  HSE0    & 1.128 \cite{22} \newline 1.110 \cite{23} & K\(\rightarrow\)K \cite{22} & up-up \cite{22} \\
         &  G0W0    & 0.30 \cite{20} \newline 1.334 \cite{23} & G\(\rightarrow\)M \newline \cite{20,23} & up-down \cite{20,23} \\ 
     \bottomrule
    \end{tabular*}
\end{table}

\begin{figure}[htbp]
\centering
\includegraphics[width=8.3cm, height=15.cm]{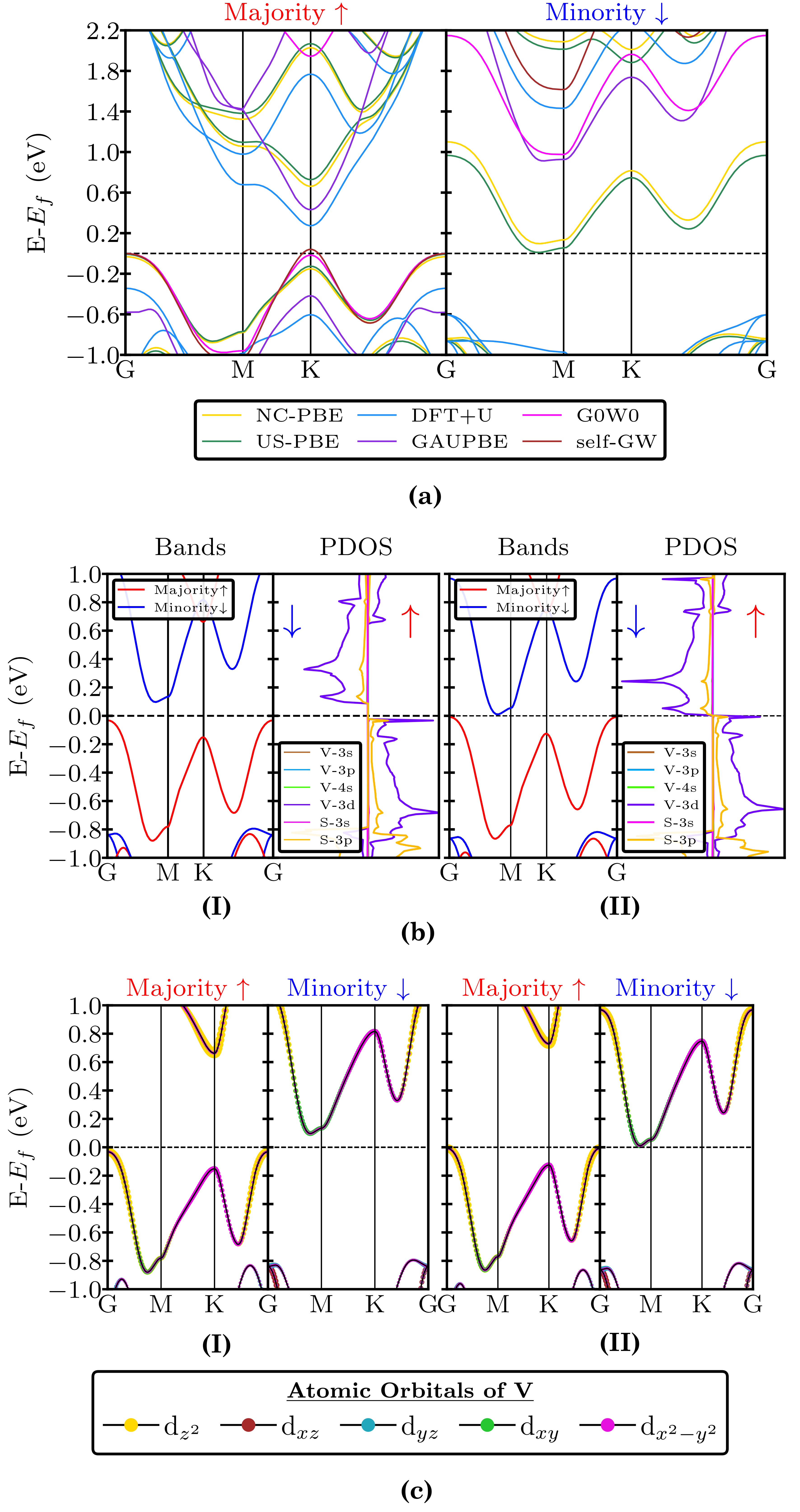}
\caption{\label{fgr:figure2}(a) Band Structure of monolayer VS$_2$ with DFT (ONCV-PBE and US-PBE), DFT+U, GAUPBE, G0W0, and self-GW, (b) band structures and projected density of states (PDOS) of ONCV-PBE (I) and US-PBE (II), (c) V atom and the d orbital decomposed fatband structures of monolayer VS$_2$, ONCV-PBE (I) and US-PBE (II).}
\end{figure}

\par Table~\ref{tbl:table2}, when considered in conjunction with Figure~\ref{fgr:figure2}(a), which illustrates the band structure of the H phase, provides a comprehensive understanding of VS$_2$'s electronic properties. Importantly, Figure~\ref{fgr:figure2}(a) highlights the prominent transition points with high symmetry, such as G, M, and K, as illustrated in Table~\ref{tbl:table2}. Methods based on ONCV PP, such as PBE, G0W0, and Self-GW, reveal that the gap energies are 0.129 eV, 0.987 eV, and 1.61 eV, respectively, all with an identical transition point from G to M, indicating an indirect band gap. The spin channel has been noted as up-down for all these methods. For US PP methods such as PBE, DFT+U, and GAUPBE, the gap values produced are 0.018 eV, 0.617 eV, and 0.946 eV, respectively. Here, the PBE method has shown an indirect band gap transition from G to M, while DFT+U and GAUPBE have indicated an indirect transition from G to K. The spin channels are predominantly up-down for PBE and up-up for both DFT+U and GAUPBE. Additionally, previous studies using the PBE method have demonstrated direct band gap values of 0.187 eV \cite{22} and 0.046 eV \cite{23} with a uniform transition point from G to M and an up-down spin channel. Conversely, the DFT+U method has primarily presented indirect band gaps, with transition points from G to M and G to K. The band gaps reported for the DFT+U method are 0.26 eV \cite{20}, 0.721 eV \cite{22}, and 0.473 eV \cite{23}, respectively, and the spin channel is always up-down except for the transition from G to K, which has an up-up spin channel. It is emphasized that the DFT+U method typically increases the bandgap value compared to the PBE method. The reason for this is that the DFT+U method takes into account the on-site Coulomb interaction between electrons, which can be significant for strongly correlated materials. The HSE0 and G0W0 methods have also shown a variety of direct \cite{22} and indirect \cite{23} bandgap transitions as seen in previous studies.

\par Figure~\ref{fgr:figure2}(b) highlights the band and partial density of states (PDOS) structure of the material derived from both ONCV-PBE and US-PBE PPs. It becomes evident that the 3d orbital of V exerts a more pronounced influence on the band structure of a monolayer of VS$_2$ than the 3p orbital of S. Furthermore, Figure~\ref{fgr:figure2}(c) specifically highlights the fatband structure associated with the d orbital of the V atom, emphasizing its significant role in the material's electronic properties. VS$_2$ possesses three degenerate states in the H phase: d$_{z^{2}}$, d$_{x^{2}-y^{2},xy}$, and d$_{xz,yz}$ \cite{24}. Figure~\ref{fgr:figure2}(c) shows that the d$_{z^{2}}$ and d$_{x^{2}-y^{2},xy}$ degenerate states have a decisive influence on the band structure of both spin channels. Concerning the valance electrons of V, the d$_{z^{2}}$ orbital is more energetic for the majority spin channel at the G high symmetry point than the d$_{xy}$ orbital at the K symmetry point. On the other hand, the d$_{x^{2}-y^{2}}$ orbital is active for the minority spin channel in the conduction band close to the M symmetry point. The results presented in Figure~\ref{fgr:figure2}(c) suggest that the indirect transition d$_{z^{2}}$ (majority spin channel) and d$_{x^{2}-y^{2}}$ (minority spin channel) orbitals occur. The results are consistent with a previous study \cite{23}. Overall, Table~\ref{tbl:table2} and Figure~\ref{fgr:figure2}(a) offers a thorough comparison of monolayer VS$_2$'s band gap results across a variety of PPs and methods, emphasizing the relevance of both direct and indirect band gap transitions. Additionally, the study conducted by Su et al. \cite{27} revealed that the semiconductor is of p-type, which aligns with our study.

\subsection{Uniaxial and Biaxial Strain Effect on Electronic Band Structure}

\begin{figure}[!b]
\centering
  \includegraphics[width=8.3cm, height=4.3cm]{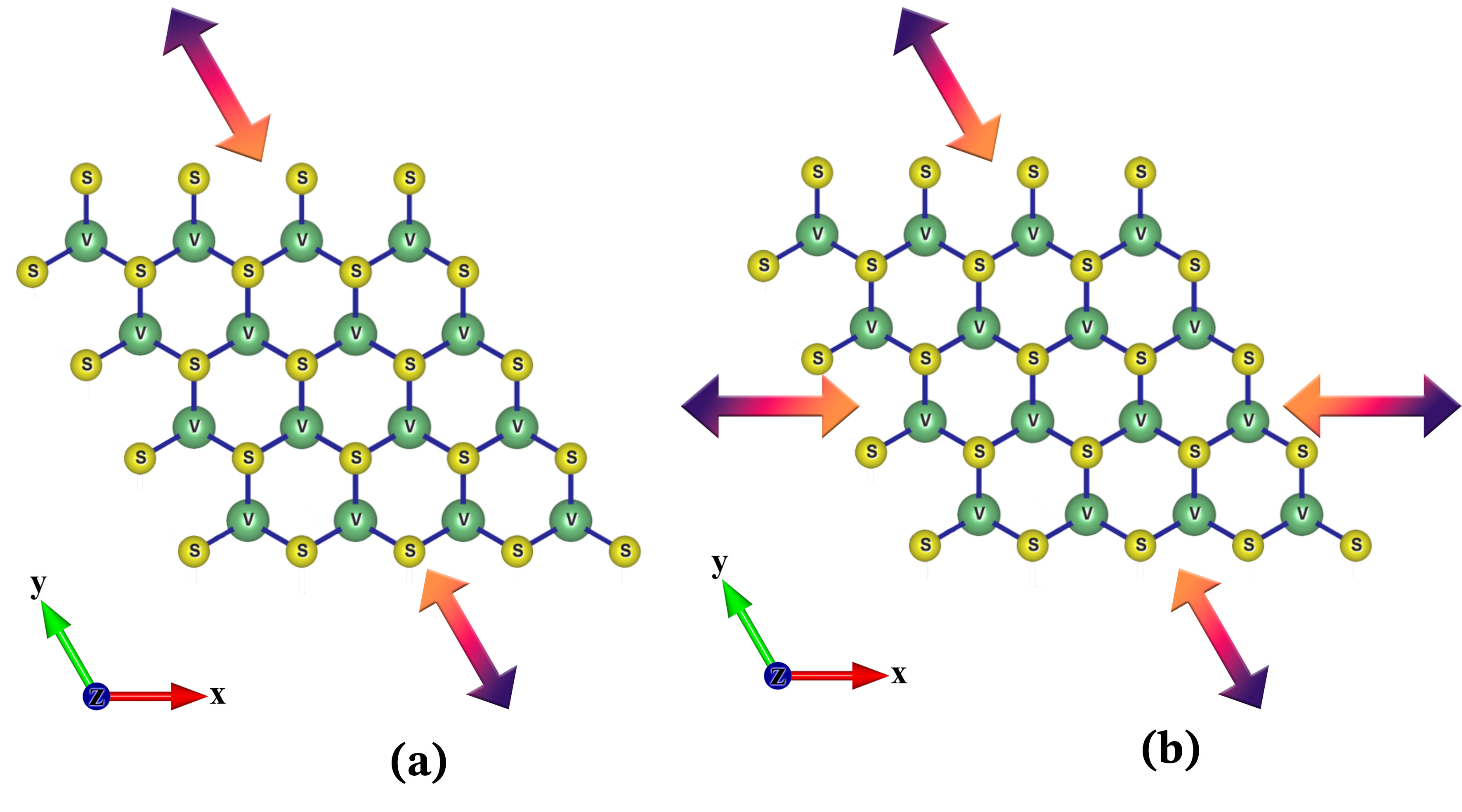}
  \caption{Lattice deformation of monolayer VS$_2$ in the H phase subjected to a strain of $\pm5\%$ (a) uniaxial, (b) biaxial along the y and x-y axes, respectively.}
  \label{fgr:figure3}
\end{figure}

 \begin{figure*}[htbp]
\centering
  \includegraphics[width=1.\textwidth, height=18cm]{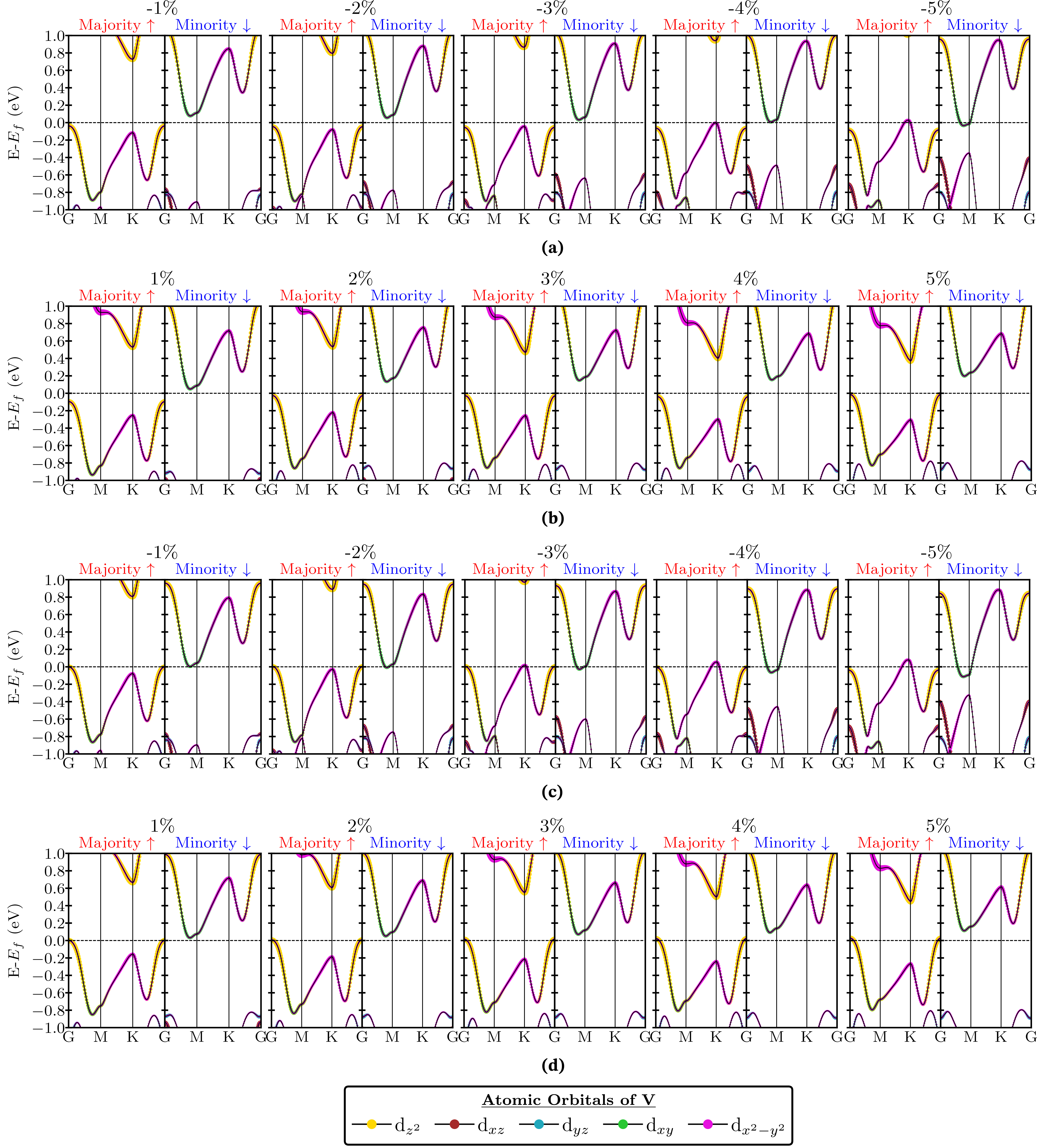}
  \caption{Fatband structure of d orbitals of monolayer VS$_2$ under uniaxial strain (a) ONCV- PBE in compressive case, (b) ONCV-PBE in tensile case, (c) US-PBE in compressive case, and (d) US-PBE in tensile case.}
  \label{fgr:figure4}
\end{figure*}

\par In 2D materials, the disruption of inversion symmetry along the z axis contrasts starkly with traditional materials, significantly influencing their electronic structures. The materials stable in the H phase possess a \textit{D$_{6h}$} group crystal symmetry and maintain inversion symmetry for H bulk and even-layered films. Conversely, structures with an odd number of layers, including monolayers, lack inversion symmetry and exhibit \textit{D$_{3h}$} point group symmetry \cite{14}. Given this symmetry alteration, distortions in the lattice parameters along the x and y axes profoundly impact the electronic properties of VS$_2$ \cite{13,22,23}. Hence, the H-phase band structure of the monolayer VS$_2$ has been investigated under deformations in its lattice parameters , subjected to both uniaxial [Figure~\ref{fgr:figure3}~(a)] and biaxial [Figure~\ref{fgr:figure3}~(b)] strains at a ratio of $\pm~5\%$.

\begin{figure*}[htbp]
\centering
  \includegraphics[width=1.\textwidth, height=18cm]{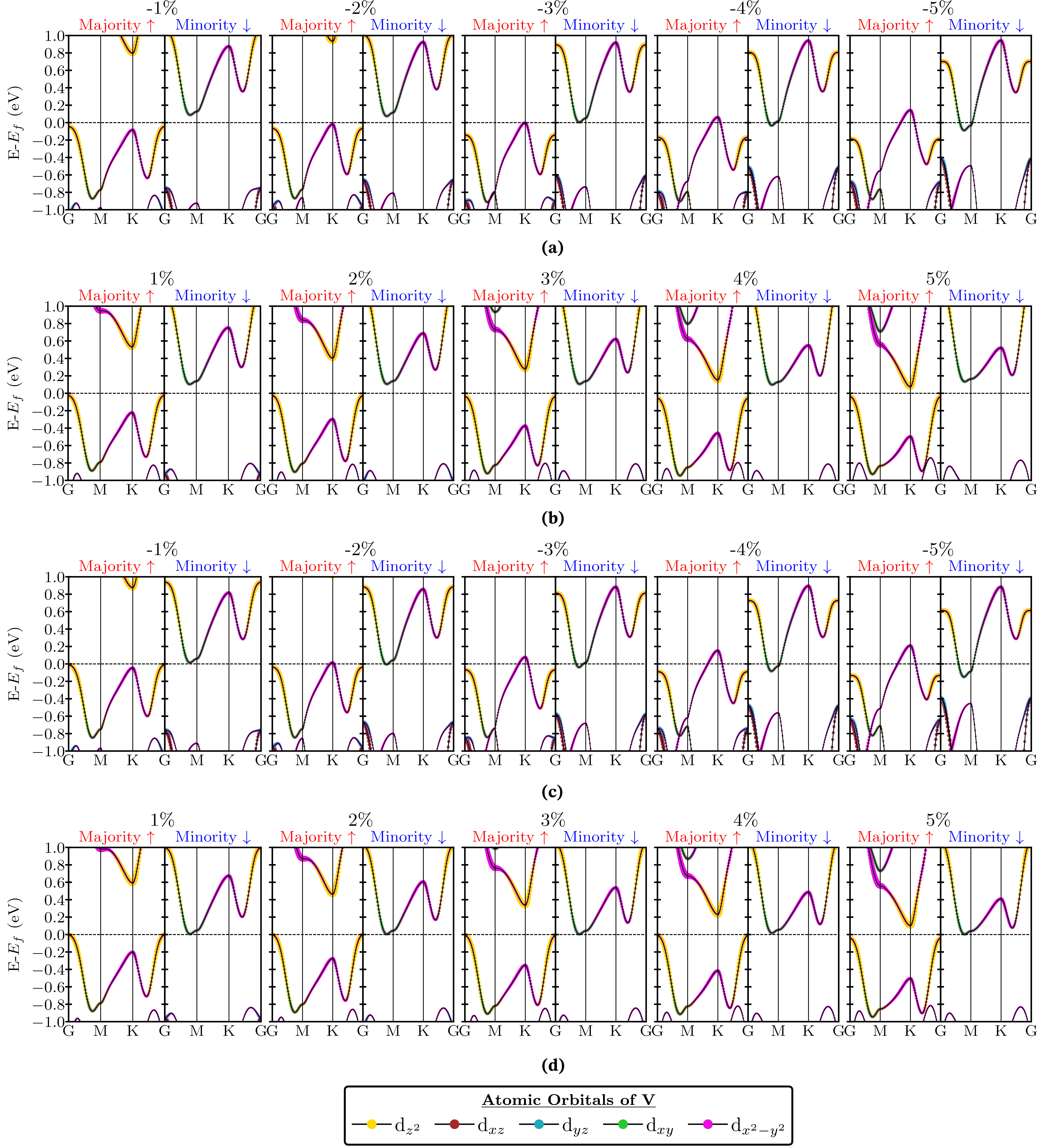}
  \caption{Fatband structure of d orbitals of monolayer VS$_2$ under biaxial strain (a) ONCV- PBE in compressive case, (b) ONCV-PBE in tensile case, (c) US-PBE in compressive case, and (d) US-PBE in tensile case.}
  \label{fgr:figure5}
\end{figure*}

\par In order to delve into the evolution of the band structure, the fatband including 3d orbitals of the V atom is examined under uniaxial (along the y-axis) and biaxial strains (along the x-y axes). Figure~\ref{fgr:figure4} demonstrates the progression of the monolayer VS$_2$ band structure under uniaxial tensile [Figure~\ref{fgr:figure4}(b)-(d)] and compressive strain [Figure~\ref{fgr:figure4}(a)-(c)]. As seen in Figure~\ref{fgr:figure4}(a)-(c), the majority spin of the d$_{xy}$ orbital approaches the Fermi level at the K point before crossing it as compressive strain increases from -1\% to -5\%. The majority spin surpasses the Fermi level at -4\% for ONCV-PBE and at -3\% for US-PBE. Furthermore, electron activity at the G point shifts to valence electrons at the K point, although this does not occur in the conduction band. In both types of PP methods, the movement of electrons in the conduction band is mainly controlled by the d$_{xy}$ and d$_{x^{2}-y^{2}}$ orbitals at the M point. It is noted that by applying compressive strain, the majority and minority spin channels intersect and achieve a metallic state. In both types of PP methods, the movement of electrons in the conduction band is mainly controlled by the d$_{xy}$ and d$_{x^{2}-y^{2}}$ orbitals when at the M point. It is noted that by applying compressive strain, majority and minority spin channels intersect and achieve a metallic state.

\par Figure~\ref{fgr:figure4}(b)-(d) presents the evolution of the band structure under uniaxial tensile strain. It indicates that monolayer VS$_2$ does not exhibit a metallic transition between spin channels compared to compressive strain. Moreover, unlike with compressive strain analysis, the major spin at the K-point (in the valence band) moves further from the Fermi level, while those in the conduction band approach it. The band gap occurs between the majority spins of d$_{z^{2}}$ and minority spin channels of d$_{xy}$ and d$_{x^{2}-y^{2}}$ (G to M). Whereas increasing tensile strain from 1\% to 5\% does not alter the orbital or high symmetry points where gaps occur, it is deduced that the band gap of monolayer VS$_2$ increases monotonically with strain. The transition observed is mechanistically attributed to the strain-induced change in atomic bond lengths within VS$_2$, which increases atomic orbital overlap \cite{44,45}. The increased overlap amplifies the density of electronic states near the Fermi energy, consequently promoting electronic transitions into the conduction band. Additionally, previous studies on MoS$_2$ \cite{46} and NbS$_2$ \cite{47} have reported that applying uniaxial strain brings about the breaking down of \textit{D$_{3h}$} point group symmetry to \textit{C$_{2v}$} group. Such symmetry breaking directly influences the electronic behavior of the d orbitals of the V atom.
\par Figure~\ref{fgr:figure5} depicts the fatband structure of d orbitals in a monolayer VS$_2$ subjected to biaxial strain. Figure~\ref{fgr:figure5}(a)-(c) reveals that, for the majority of spin channels, the valence band approaches the Fermi level, with the d$_{xy}$ orbital at the K point overtaking the d$_{z^{2}}$ orbital at point G. The band gap between majority and minority spins is nearly zero under ONCV-PBE -3\% and US-PBE -2\% compressive strains. As the strain increases, the material adopts a metallic nature, as observed in uniaxial strain [Figure~\ref{fgr:figure4}(a)-(c)]. However, the metallic property appears at a lower strain level in biaxial strain compared to uniaxial strain.

\par In the case of biaxial tensile strain, the gap energy remains relatively unchanged in monolayer VS$_2$. Additionally, the gap arises between majority spins of d$_{z^{2}}$ at point G and minority spin channels of d$_{xy}$ and d$_{x^{2}-y^{2}}$ at point M, similar to uniaxial strain cases. The biaxial strain does not impact electrons situated at higher symmetry points G and M but directly influences those located at point K. As depicted in Figure~\ref{fgr:figure5}(b)-(d), there is a noticeable shift of d$_{z^{2}}$ electrons in the conduction band toward the Fermi level as the strain increases from 1\% to 5\%. Notably, only in ONCV-PBE, at 5\% strain, do the d$_{xy}$ and d$_{x^{2}-y^{2}}$ orbitals at M (in the minority spin channel) get replaced by the d$_{z^{2}}$ orbital at K on the conduction level. The electronic phase transition in the biaxial strain case is directly dependent on the bond length between V and S atoms since the symmetry elements of the system do not alter.

\begin{figure}[t]
\centering
  \includegraphics[width=8.2cm, height=6.2cm]{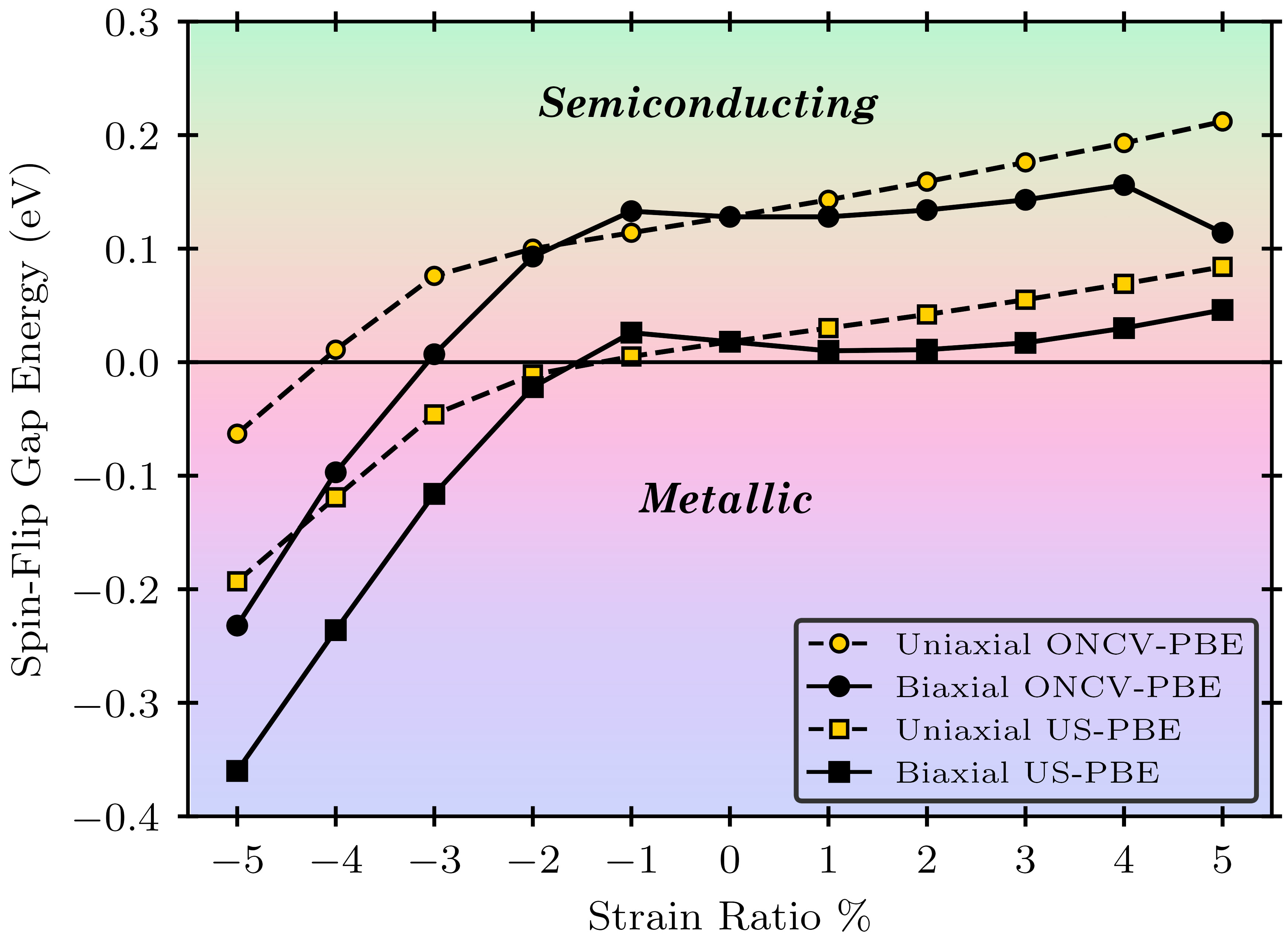}
  \caption{Evolution of the spin-flip gap energy of monolayer VS$_2$ with strain.}
  \label{fgr:figure6}
\end{figure}

\par Based on the above results, it is concluded that the energy gap in monolayer VS$_2$ exists between different spin channels, a phenomenon referred to as spin-flip energy. Figure~\ref{fgr:figure6} illustrates the evolution of the spin-flip gap energy versus strain ratio for both uniaxial and biaxial strains. The uniaxial compressive strain leads to a transition from semiconductor to spin-gapless semiconductor (type II) \cite{48,49} in monolayer VS$_2$. As the strain ratio increases, the semiconductor electronic phase eventually transforms into the metallic phase. In contrast, in the uniaxial tensile strain case, no electronic phase transition occurs as in the compressive strain case; however, the spin-flip gap energy increases monotonically with the strain ratio. Similarly, in the biaxial case, when subjected to biaxial compressive strain, the VS$_2$ monolayer undergoes an electronic phase transformation. Differing from the uniaxial strain case, the spin-gapless (type II) phase emerges at -3\% and -2\% for ONCV-PBE and US-PBE, respectively, and then the metallic phase becomes apparent below these strain ratios. It is found that biaxial tensile strain does not trigger an electronic phase transition nor affects the spin-flip gap energy as in the uniaxial strain cases. The influence of biaxial strain on the band structure of the material, as previously studied in ab initio studies \cite{22,23}, is remarkably consistent with our study. Normally, the gap observed between opposite spin channels exists primarily within the same spin channels when subjected to a 5\% biaxial tensile strain in Figure~\ref{fgr:figure5}(b). It is assumed that this discrepancy is due to the ONCV PP.

\par Overall, these results highlight that biaxial compressive strain induces electronic phase transitions in monolayer VS$_2$ at lower strain rates compared to uniaxial strains. Nonetheless, when it comes to tensile strains, uniaxial deformation has a more significant impact on altering spin-flip gap energy than its biaxial counterpart.

\section{Conclusions}
\par The electronic structure of monolayer VS$_2$ has been extensively examined using two types of pseudopotentials, namely ONCV and US. The selection of ONCV for its precision and US for its computational efficiency was crucial in analyzing VS$_2$'s strain-induced electronic transitions. This selection highlights the significance of pseudopotential choice in studying the electronic properties of 2D materials.

\par Various computational methods, including PBE, DFT+U, GAUPBE, G0W0, and self-GW, were employed to determine the band gap of monolayer VS$_2$. The calculated band gap values were 0.129, 0.018, 0.617, 0.946, 0.987, and 1.610 eV, respectively. It is noteworthy that a gap between different spin channels exists, with the exception of DFT+U and GAUPBE.

\par Our research reveals that monolayer VS$_2$ undergoes a phase shift from a semiconducting to a metallic state when subjected to both uniaxial and biaxial compressive strains, where the biaxial strain corresponds to hydrostatic pressure. However, the transition to metallization occurs at lower levels of biaxial strain, due to the more uniform stress distribution compared to uniaxial strain. Additionally, biaxial strain's simultaneous decrease in bond lengths along both x and y lattice vectors leads to a more pronounced orbital overlap at lower strain values, facilitating an earlier onset of metallization. This emphasizes the distinct impact of strain types on the electronic structure of monolayer VS$_2$. While uniaxial tensile strain monotonically increases the gap energy, biaxial tensile strain does not have a similar impact, highlighting the unique strain-dependent electronic properties of VS$_2$.

\par In summary, our results show that biaxial strain is more effective in inducing metallization in monolayer VS$_2$ than uniaxial strain. This advanced understanding of different strain behaviors is critical for the design of advanced electronic devices, including heterojunctions and straintronic applications.

\section{Acknowledgement}

This work was supported by the Management Unit of Scientific Research Projects of Firat University (FÜBAP)(Project Number: FF.16.28). Ş.Ö. is supported by TÜBİTAK under grant no. 120F354. Computing resources used in this work were provided by the National Center for High Performance Computing (UHeM) under grant no. 1007872020 and TUBİTAK ULAKBİM, High Performance and Grid Computing Center (TRUBA resources).


\bibliographystyle{unsrt}
\bibliography{references}  

\begin{thebibliography}{10}

\bibitem{1}
Qing~Hua Wang, Kourosh Kalantar-Zadeh, Andras Kis, Jonathan~N. Coleman, and Michael~S. Strano.
\newblock {Electronics and optoelectronics of two-dimensional transition metal dichalcogenides}.
\newblock {\em Nature Nanotechnology}, 7(11):699--712, 2012.

\bibitem{2}
Sajedeh Manzeli, Dmitry Ovchinnikov, Diego Pasquier, Oleg~V. Yazyev, and Andras Kis.
\newblock {2D transition metal dichalcogenides}.
\newblock {\em Nature Reviews Materials}, 2(8):1--15, 2017.

\bibitem{3}
Manish Chhowalla, Hyeon~Suk Shin, Goki Eda, Lain-Jong Li, Kian~Ping Loh, and Hua Zhang.
\newblock {The chemistry of two-dimensional layered transition metal dichalcogenide nanosheets}.
\newblock {\em Nature Chemistry}, 5(4):263--275, 2013.

\bibitem{4}
John~R. Schaibley, Hongyi Yu, Genevieve Clark, Pasqual Rivera, Jason~S. Ross, Kyle~L. Seyler, Wang Yao, and Xiaodong Xu.
\newblock {Valleytronics in 2D materials}.
\newblock {\em Nature Reviews Materials}, 1(11):16055, 2016.

\bibitem{5}
Narayanasamy~Sabari Arul and Vellalapalayam~Devaraj Nithya.
\newblock {\em Two Dimensional Transition Metal Dichalcogenides}.
\newblock Springer Singapore, Singapore, 2019.

\bibitem{6}
Nicholas~R. Glavin, Rahul Rao, Vikas Varshney, Elisabeth Bianco, Amey Apte, Ajit Roy, Emilie Ringe, and Pulickel~M. Ajayan.
\newblock {Emerging Applications of Elemental 2D Materials}.
\newblock {\em Advanced Materials}, 32(7):1--22, 2020.

\bibitem{7}
Xiaofei Zhou, Hainan Sun, and Xue Bai.
\newblock {Two-Dimensional Transition Metal Dichalcogenides: Synthesis, Biomedical Applications and Biosafety Evaluation}.
\newblock {\em Frontiers in Bioengineering and Biotechnology}, 8(April):1--13, 2020.

\bibitem{8}
Jianwei Su, Guiheng Liu, Lixin Liu, Jiazhen Chen, Xiaozong Hu, Yuan Li, Huiqiao Li, and Tianyou Zhai.
\newblock {Recent Advances in 2D Group VB Transition Metal Chalcogenides}.
\newblock {\em Small}, 17(14):2005411, 2021.

\bibitem{9}
Heejun Yang, Sung~Wng Kim, Manish Chhowalla, and Young~Hee Lee.
\newblock {Structural and quantum-state phase transitions in van der Waals layered materials}.
\newblock {\em Nature Physics}, 13(10):931--937, 2017.

\bibitem{10}
Xinmao Yin, Chi~Sin Tang, Yue Zheng, Jing Gao, Jing Wu, Hua Zhang, Manish Chhowalla, Wei Chen, and Andrew~T.S. Wee.
\newblock {Recent developments in 2D transition metal dichalcogenides: Phase transition and applications of the (quasi-)metallic phases}.
\newblock {\em Chemical Society Reviews}, 50(18):10087--10115, 2021.

\bibitem{11}
Wenbin Li, Xiaofeng Qian, and Ju~Li.
\newblock {Phase transitions in 2D materials}.
\newblock {\em Nature Reviews Materials}, 6(9):829--846, 2021.

\bibitem{12}
Junyao Li, Xingxing Li, and Jinlong Yang.
\newblock {A review of bipolar magnetic semiconductors from theoretical aspects}.
\newblock {\em Fundamental Research}, 2(4):511--521, 2022.

\bibitem{13}
Huei-Ru Fuh, Binghai Yan, Shu-Chun Wu, Claudia Felser, and Ching-Ray Chang.
\newblock {Metal-insulator transition and the anomalous Hall effect in the layered magnetic materials VS 2 and VSe 2}.
\newblock {\em New Journal of Physics}, 18(11):113038, 2016.

\bibitem{14}
Gui~Bin Liu, Di~Xiao, Yugui Yao, Xiaodong Xu, and Wang Yao.
\newblock {Electronic structures and theoretical modelling of two-dimensional group-VIB transition metal dichalcogenides}.
\newblock {\em Chemical Society Reviews}, 44(9):2643--2663, 2015.

\bibitem{15}
Rusen Yan, Guru Khalsa, Brian~T. Schaefer, Alexander Jarjour, Sergei Rouvimov, Katja~C. Nowack, Huili~G. Xing, and Debdeep Jena.
\newblock {Thickness dependence of superconductivity in ultrathin NbS 2}.
\newblock {\em Applied Physics Express}, 12(2):023008, 2019.

\bibitem{16}
Avinash~P. Nayak, Swastibrata Bhattacharyya, Jie Zhu, Jin Liu, Xiang Wu, Tribhuwan Pandey, Changqing Jin, Abhishek~K. Singh, Deji Akinwande, and Jung~Fu Lin.
\newblock {Pressure-induced semiconducting to metallic transition in multilayered molybdenum disulphide}.
\newblock {\em Nature Communications}, 5(May):1--9, 2014.

\bibitem{17}
Seunghyun Song, Dong~Hoon Keum, Suyeon Cho, David Perello, Yunseok Kim, and Young~Hee Lee.
\newblock {Room Temperature Semiconductor-Metal Transition of MoTe2 Thin Films Engineered by Strain}.
\newblock {\em Nano Letters}, 16(1):188--193, 2016.

\bibitem{18}
Wenhui Hou, Ahmad Azizimanesh, Arfan Sewaket, Tara Pe{\~{n}}a, Carla Watson, Ming Liu, Hesam Askari, and Stephen~M. Wu.
\newblock {Strain-based room-temperature non-volatile MoTe2 ferroelectric phase change transistor}.
\newblock {\em Nature Nanotechnology}, 14(7):668--673, 2019.

\bibitem{19}
Md~Rasidul Islam, Md~Rayid~Hasan Mojumder, Biazid~Kabir Moghal, A.~S.M.Jannatul Islam, Mohammad~Raza Miah, Sourav Roy, Anuj Kumar, A.~S.M. Shihavuddin, and Ratil~H. Ashique.
\newblock {Impact of strain on the electronic, phonon, and optical properties of monolayer transition metal dichalcogenides XTe2(X = Mo and W)}.
\newblock {\em Physica Scripta}, 97(4):45806, 2022.

\bibitem{20}
Nannan Luo, Chen Si, and Wenhui Duan.
\newblock {Structural and electronic phase transitions in ferromagnetic monolayer VS2 induced by charge doping}.
\newblock {\em Physical Review B}, 95(20):205432, 2017.

\bibitem{21}
Yuanyuan Cui, Wei Fan, Yujie Ren, Guang Yang, and Yanfeng Gao.
\newblock {First-principles calculations to study the optical/electronic properties of 2D VS2 with Z doping (Z = N, P, As, F, Cl and Br)}.
\newblock {\em Progress in Natural Science: Materials International}, 32(2):236--241, 2022.

\bibitem{22}
Min Kan, Bo~Wang, Young~Hee Lee, and Qiang Sun.
\newblock {A density functional theory study of the tunable structure, magnetism and metal-insulator phase transition in VS2 monolayers induced by in-plane biaxial strain}.
\newblock {\em Nano Research}, 8(4):1348--1356, 2015.

\bibitem{23}
Huei-Ru Fuh, Ching-Ray Chang, Yin-Kuo Wang, Richard F.~L. Evans, Roy~W. Chantrell, and Horng-Tay Jeng.
\newblock {Newtype single-layer magnetic semiconductor in transition-metal dichalcogenides VX2 (X = S, Se and Te)}.
\newblock {\em Scientific Reports}, 6(1):32625, 2016.

\bibitem{24}
D.~Sahoo, S.~Senapati, and R.~Naik.
\newblock {Progress and prospects of 2D VS2 transition metal dichalcogenides}.
\newblock {\em FlatChem}, 36(October):100455, 2022.

\bibitem{25}
Camiel van Efferen, Jan Berges, Joshua Hall, Erik van Loon, Stefan Kraus, Arne Schobert, Tobias Wekking, Felix Huttmann, Eline Plaar, Nico Rothenbach, Katharina Ollefs, Lucas~Machado Arruda, Nick Brookes, Gunnar Sch{\"{o}}nhoff, Kurt Kummer, Heiko Wende, Tim Wehling, and Thomas Michely.
\newblock {A full gap above the Fermi level: the charge density wave of monolayer VS2}.
\newblock {\em Nature Communications}, 12(1):1--9, 2021.

\bibitem{26}
Tappei Kawakami, Katsuaki Sugawara, Hirofumi Oka, Kosuke Nakayama, Ken Yaegashi, Seigo Souma, Takashi Takahashi, Tomoteru Fukumura, and Takafumi Sato.
\newblock {Charge-density wave associated with higher-order Fermi-surface nesting in monolayer VS2}.
\newblock {\em npj 2D Materials and Applications}, 7(1):35, 2023.

\bibitem{27}
Jianwei Su, Mingshan Wang, Yuan Li, Fakun Wang, Qiao Chen, Peng Luo, Junbo Han, Shun Wang, Huiqiao Li, and Tianyou Zhai.
\newblock {Sub-Millimeter-Scale Monolayer p-Type H‐Phase VS 2}.
\newblock {\em Advanced Functional Materials}, 30(17):2000240, 2020.

\bibitem{tariq2021pristine}
Zeeshan Tariq, Sajid~Ur Rehman, Xiaoming Zhang, Faheem~K Butt, Shuai Feng, Bakhtiar~Ul Haq, Jun Zheng, Buwen Cheng, and Chuanbo Li.
\newblock Pristine and janus monolayers of vanadium dichalcogenides: potential materials for overall water splitting and solar energy conversion.
\newblock {\em Journal of Materials Science}, 56(21):12270--12284, 2021.

\bibitem{28}
Feng Miao, Shi~Jun Liang, and Bin Cheng.
\newblock {Straintronics with van der Waals materials}.
\newblock {\em npj Quantum Materials}, 6(1):2--5, 2021.

\bibitem{kim2021electronic}
Hyuk~Jin Kim, Byoung~Ki Choi, In~Hak Lee, Min~Jay Kim, Seung-Hyun Chun, Chris Jozwiak, Aaron Bostwick, Eli Rotenberg, and Young~Jun Chang.
\newblock Electronic structure and charge-density wave transition in monolayer vs2.
\newblock {\em Current Applied Physics}, 30:8--13, 2021.

\bibitem{29}
Shyue~Ping Ong, William~Davidson Richards, Anubhav Jain, Geoffroy Hautier, Michael Kocher, Shreyas Cholia, Dan Gunter, Vincent~L. Chevrier, Kristin~A. Persson, and Gerbrand Ceder.
\newblock {Python Materials Genomics (pymatgen): A robust, open-source python library for materials analysis}.
\newblock {\em Computational Materials Science}, 68:314--319, 2013.

\bibitem{30}
Paolo Giannozzi, Stefano Baroni, Nicola Bonini, Matteo Calandra, Roberto Car, Carlo Cavazzoni, Davide Ceresoli, Guido~L. Chiarotti, Matteo Cococcioni, Ismaila Dabo, Andrea {Dal Corso}, Stefano de~Gironcoli, Stefano Fabris, Guido Fratesi, Ralph Gebauer, Uwe Gerstmann, Christos Gougoussis, Anton Kokalj, Michele Lazzeri, Layla Martin-Samos, Nicola Marzari, Francesco Mauri, Riccardo Mazzarello, Stefano Paolini, Alfredo Pasquarello, Lorenzo Paulatto, Carlo Sbraccia, Sandro Scandolo, Gabriele Sclauzero, Ari~P. Seitsonen, Alexander Smogunov, Paolo Umari, and Renata~M. Wentzcovitch.
\newblock {QUANTUM ESPRESSO: a modular and open-source software project for quantum simulations of materials}.
\newblock {\em Journal of Physics: Condensed Matter}, 21(39):395502, 2009.

\bibitem{31}
Gianluca Prandini, Antimo Marrazzo, Ivano~E. Castelli, Nicolas Mounet, and Nicola Marzari.
\newblock {Precision and efficiency in solid-state pseudopotential calculations}.
\newblock {\em npj Computational Materials}, 4(1):72, 2018.

\bibitem{32}
David Vanderbilt.
\newblock {Soft self-consistent pseudopotentials in a generalized eigenvalue formalism}.
\newblock {\em Physical Review B}, 41(11):7892--7895, 1990.

\bibitem{33}
Kevin~F. Garrity, Joseph~W. Bennett, Karin~M. Rabe, and David Vanderbilt.
\newblock {Pseudopotentials for high-throughput DFT calculations}.
\newblock {\em Computational Materials Science}, 81:446--452, 2014.

\bibitem{34}
D.~R. Hamann.
\newblock {Optimized norm-conserving Vanderbilt pseudopotentials}.
\newblock {\em Physical Review B - Condensed Matter and Materials Physics}, 88(8):1--10, 2013.

\bibitem{schlipf2015optimization}
Martin Schlipf and Fran{\c{c}}ois Gygi.
\newblock Optimization algorithm for the generation of oncv pseudopotentials.
\newblock {\em Computer Physics Communications}, 196:36--44, 2015.

\bibitem{35}
John~P. Perdew, Kieron Burke, and Matthias Ernzerhof.
\newblock {Generalized gradient approximation made simple}.
\newblock {\em Physical Review Letters}, 77(18):3865--3868, 1996.

\bibitem{36}
Jong~Won Song, Giacomo Giorgi, Koichi Yamashita, and Kimihiko Hirao.
\newblock {Communication: Singularity-free hybrid functional with a Gaussian-attenuating exact exchange in a plane-wave basis}.
\newblock {\em Journal of Chemical Physics}, 138(24), 2013.

\bibitem{37}
Joseph~W. Bennett, Blake~G. Hudson, Irene~K. Metz, Dongyue Liang, Sidney Spurgeon, Qiang Cui, and Sara~E. Mason.
\newblock {A systematic determination of hubbard U using the GBRV ultrasoft pseudopotential set}.
\newblock {\em Computational Materials Science}, 170(August):109137, 2019.

\bibitem{38}
Giovanni Pizzi, Valerio Vitale, Ryotaro Arita, Stefan Bl{\"u}gel, Frank Freimuth, Guillaume G{\'e}ranton, Marco Gibertini, Dominik Gresch, Charles Johnson, Takashi Koretsune, et~al.
\newblock {Wannier90 as a community code: New features and applications}.
\newblock {\em Journal of Physics Condensed Matter}, 32(16):165902, 2020.

\bibitem{39}
D.~Sangalli, A.~Ferretti, H.~Miranda, C.~Attaccalite, I.~Marri, E.~Cannuccia, P.~Melo, M.~Marsili, F.~Paleari, A.~Marrazzo, G.~Prandini, P.~Bonf{\`{a}}, M.~O. Atambo, F.~Affinito, M.~Palummo, A.~Molina-S{\'{a}}nchez, C.~Hogan, M.~Gr{\"{u}}ning, D.~Varsano, and A.~Marini.
\newblock {Many-body perturbation theory calculations using the yambo code}.
\newblock {\em Journal of Physics Condensed Matter}, 31(32):325902, 2019.

\bibitem{40}
Mark~S. Hybertsen and Steven~G. Louie.
\newblock {Electron correlation in semiconductors and insulators: Band gaps and quasiparticle energies}.
\newblock {\em Physical Review B}, 34(8):5390--5413, 1986.

\bibitem{41}
Wolfgang von~der Linden and Peter Horsch.
\newblock {Precise quasiparticle energies and Hartree-Fock bands of semiconductors and insulators}.
\newblock {\em Physical Review B}, 37(14):8351--8362, 1988.

\bibitem{42}
Alberto Guandalini, Pino D’Amico, Andrea Ferretti, and Daniele Varsano.
\newblock Efficient gw calculations in two dimensional materials through a stochastic integration of the screened potential.
\newblock {\em npj Computational Materials}, 9(1):44, 2023.

\bibitem{43}
Yandong Ma, Ying Dai, Meng Guo, Chengwang Niu, Yingtao Zhu, and Baibiao Huang.
\newblock {Evidence of the Existence of Magnetism in Pristine VX 2 Monolayers (X = S, Se) and Their Strain-Induced Tunable Magnetic Properties}.
\newblock {\em ACS Nano}, 6(2):1695--1701, 2012.

\bibitem{44}
Luqing Wang, Alex Kutana, and Boris~I. Yakobson.
\newblock {Many-body and spin-orbit effects on direct-indirect band gap transition of strained monolayer MoS 2 and WS 2}.
\newblock {\em Annalen der Physik}, 526(9-10):L7--L12, 2014.

\bibitem{45}
Yangwu Wu, Lu~Wang, Huimin Li, Qizhi Dong, and Song Liu.
\newblock {Strain of 2D materials via substrate engineering}.
\newblock {\em Chinese Chemical Letters}, 33(1):153--162, 2022.

\bibitem{46}
G~Kukucska and J~Koltai.
\newblock {Theoretical Investigation of Strain and Doping on the Raman Spectra of Monolayer MoS 2}.
\newblock {\em physica status solidi (b)}, 254(11):1--5, 2017.

\bibitem{47}
Rui-Chun Xiao, Ding-Fu Shao, Zhi-Qiang Zhang, and Hua Jiang.
\newblock {Two-Dimensional Metals for Piezoelectriclike Devices Based on Berry-Curvature Dipole}.
\newblock {\em Physical Review Applied}, 13(4):044014, 2020.

\bibitem{48}
Zengji Yue, Zhi Li, Lina Sang, and Xiaolin Wang.
\newblock {Spin-Gapless Semiconductors}.
\newblock {\em Small}, 16(31):16, 2020.

\bibitem{49}
Deepika Rani, Lakhan Bainsla, Aftab Alam, and K.~G. Suresh.
\newblock {Spin-gapless semiconductors: Fundamental and applied aspects}.
\newblock {\em Journal of Applied Physics}, 128(22):220902, 2020.

\end{thebibliography}






\end{document}